\documentstyle[prl,aps,multicol,psfig]{revtex}
\begin{document}
\draft
\hyphenation{following}

\title{Symmetry Analysis of the Nonlinear Optical Response: 
New Magneto-Optical Effect in Second Harmonic Generation}
\author{A. D\"ahn, W. H\"ubner, and K. H. Bennemann}
\address{Institute for Theoretical Physics, Freie Universit\"at 
Berlin, Arnimallee 14, D-14195 Berlin, Germany}
\date{\today}
\maketitle
\begin{abstract}
Using symmetry arguments we show how optical second harmonic generation (SHG) 
can be used to detect 
antiferromagnetism at surfaces and in thin films. Based 
on the group theoretical analysis of the nonlinear electric susceptibility
 we propose a 
new nonlinear magneto-optical effect, which allows even in the presence of 
unit-cell doubling for the unambiguous discrimination of antiferromagnetic 
surface spin configurations from ferro- or paramagnetic ones. As an example
 for this effect we discuss the polarization dependence of
 SHG from the fcc (001) surface of NiO in some detail. 
\end{abstract}
\begin{multicols}{2}
 In the last decade  optical second-harmonic generation (SHG) has become a 
well-established probe for the investigation of the geometrical and 
electronic properties of surfaces and interfaces. In particular the 
nonlinear magneto-optical Kerr effect (NOLIMOKE) has been demonstrated 
both theoretically and experimentally as a powerful technique for the 
study of surface and interface ferromagnetism due to
 its enhanced surface sensitivity \cite{pan}-\cite{rau}. To 
generalize NOLIMOKE to arbitrary spin configurations and, since under 
ambient conditions many transition metals tend to form antiferromagnetic 
surface oxides (which become more and more important as substrates 
for catalytic reactions), it is of 
considerable interest to explore the possibility of sensing not only
 bulk \cite{ere}-\cite{krich} but also surface and 
interface antiferromagnetism by SHG. Here we show by 
extending our previous theoretical work that also antiferromagnetic 
surfaces can be clearly detected by SHG. At first sight this seems 
surprising, since both the paramagnetic and the antiferromagnetic state 
exhibit time reversal symmetry. However, as will be shown by our detailed 
symmetry classification of the nonlinear electric susceptibility tensor we 
obtain characteristic differences of the susceptibilities in the para-, 
ferro-, and antiferromagnetic phases, which then can be detected by using 
different polarizations of the incoming light. Our analysis proposed here 
is different from the recent pioneering 
studies of antiferromagnetic bulk domains of $Cr_2O_3$ by 
Fiebig {\em et al.}~\cite{fieb,froh}:
 They observed antiferromagnetic 180$^{\circ}$ 
domains of bulk $Cr_2O_3$ from a strong intensity contrast in SHG 
transmission and explained this effect by the interference 
of the nonlinear electric susceptibility (nonzero only 
in the antiferromagnetic phase) and
 magnetic susceptibility (allowed in both phases) \cite{bem}. While the 
phenomenon in $Cr_2O_3$
 results from the simultaneous action of spin-orbit coupling and trigonal
 crystal distortion \cite{gros}, we show here that SHG may also probe 
antiferromagnetism at surfaces of cubic crystals without involving 
interference effects or crystal distortions.

We start our theory from the symmetry analysis of the nonlinear susceptibility
 using classical electrodynamics. The nonlinear response 
is described by the nonlinear electrical polarization 
$P^{(2\omega)}_{el}$ acting as a source term in the wave equation and 
related to the incident photon field 
by the nonlinear electrical susceptibility $\chi^{(2\omega)}_{el}$ :
\begin{equation}
P^{(2\omega)}_{el}=\chi^{(2\omega)}_{el}:E^{(\omega)}E^{(\omega)} .
\end{equation}
Here, $\chi^{(2\omega)}_{el}$ is the nonlinear electrical susceptibility 
in dipole approximation. It is a polar tensor of rank three which is
 nonzero only at the surface of a 
cubic material. 
According to Neumann's principle this susceptibility  (which is a 
property tensor) 
has to remain invariant under any symmetry transformation
 $\mathit{l_{n,n'}} (n = i,j,k, n'= i',j',k',)$ leaving the 
 lattice invariant ~\cite{bir,bha}:
\begin{equation}
\chi^{(2\omega)}_{el,i'j'k'}=l_{i'i}l_{j'j}l_{k'k} 
\chi^{(2\omega)}_{el,ijk},\,\,
\,\,
  i,j,k=x,y,z .
\end{equation}
If time inversion R : $t\rightarrow-t$ alone or in
 combination with any space operation $\mathit{l}$ belongs to the classifying 
symmetry elements, Eq. (2) must be replaced by:
\begin{equation}
\chi^{(2\omega)}_{el,i'j'k'}=\pm l_{i'i}l_{j'j}l_{k'k} 
\chi^{(2\omega)}_{el,ijk},
\,\,\,
  i,j,k=x,y,z ,
\end{equation}
where -- (+) refers to the case when $\chi^{(2\omega)}$ changes sign under
 time-inversion R, denoted as c-tensor (remains
 invariant, i-tensor) \cite{bir}. 

Using now the symmetry transformations for a paramagnetic, ferromagnetic, 
and antiferromagnetic state we can immediately determine the resulting 
non-vanishing elements of $ \chi^{(2\omega)}_{el,ijk}$. Since,
 depending on the 
magnetic state, different tensor elements vanish, it is possible to detect 
optically antiferromagnetism by varying the polarization of the incoming 
light. The allowed tensor elements resulting from Eqs. (2) and (3) are given 
in Tab. \ref{tab1}.

 The new predicted nonlinear
 magneto-optical effect 
results from the fact that the  point groups obtained for antiferromagnetic 
configurations 
are different from the ones 
describing para- or ferromagnetic states of the same surfaces, since all 
three phases have characteristically different symmetry features: A 
ferromagnetic surface never exhibits time inversion symmetry R,
 but R may occur in 
combination with spatial symmetry operations, (i.e. the symmetry group is 
magnetic.) In the paramagnetic state, however, a surface always remains 
invariant under time reversal. Hence, for surfaces of cubic crystals 
the paramagnetic state exhibits the 
same time inversion symmetry as the antiferromagnetic state.
 But the two states usually differ in the allowed space transformations, 
because 
in the paramagnetic phase no spin configuration has to be regarded.
 
 In order to demonstrate that SHG yields a new effect, which 
sensitively probes antiferromagnetic surface spin configurations, we proceed 
as follows: (i.) We determine the magnetic space group of 
the surface lattice including the spin configuration. We find that all 
examined antiferromagnetic (001), (110) and (111) surface configurations 
are highly 
symmetric: There always exists a translation T so that time reversal R 
becomes a symmetry element of the lattice in combination with this translation.
(ii.) Since the nonzero elements of the property tensors rigorously follow 
from the point groups of a crystal, we restrict ourselves to the corresponding 
operations replacing all translations by the identity ~\cite{bir}. Thus, 
time-inversion R remains a symmetry element of all examined antiferromagnetic 
spin
 configurations. Consequently unit-cell doubling occurring for most of 
the antiferromagnetic spin configurations does not affect the 
classification. (iii.) R has to be excluded from the analysis, 
since SHG is a 
dynamical process with a preferred direction of time \cite{gros}. In this 
case Neumann's principle is restricted to pure space symmetry operations of
 \cite{bir,kom}. The
 resulting point group of space-transformations, which usually is just a 
subgroup of the classical point group characterizing the 
symmetry of the three-dimensional lattice, is used to calculate 
the nonzero tensor elements of $\chi^{(2\omega)}_{el}$ in accordance with 
Eq. (2). For these subgroups, in contrast to the three-dimensional point 
groups 
\cite{bir,bha}, the tensor elements of 
$\chi^{(2\omega)}_{el}$ have not been derived previously.
 
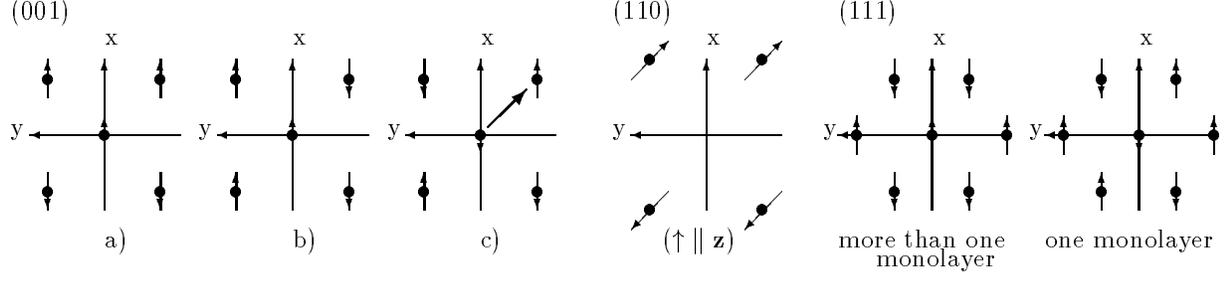
\begin{figure}
\caption{\label{fig1}Antiferromagnetic spin configurations of low 
index surfaces of NiO.}
\end{figure}

To show the predicted sensitivity of SHG to surface 
antiferromagnetism we restrict ourselves to the very 
clear-cut example of one 
prototypic spin structure for each of the fcc (001), (110) and (111) 
surfaces \cite{zus}. 
(see Tab. \ref{tab1})
In all antiferromagnetic 
configurations additional tensor elements compared to the 
paramagnetic phase appear.
As a prototypic  example how nonlinear magneto-optics detects 
antiferromagnetism unambiguously, we discuss the fcc (001) surface 
with spin configuration c)(see Fig. \ref{fig1}), which is similar to the one of 
(001) NiO \cite{nio}.
The point group describing the symmetry of this configuration consists of only
 one independent spatial symmetry operation, i.e. the 180$^{\circ}$ 
rotation about the $\mathit{z}$-axis $2_z$ ($\mathit{z}$-axis perpendicular 
to the surface, 
$\mathit{x}$- and $\mathit{y}$-axes in the surface plane), which transforms  
$ (x,y,z)\longrightarrow (-x,-y,z)$ and causes the spin-moments, parallel or 
antiparallel to the $\mathit{x}$-direction, to flip. But this spin flip, 
corresponding 
to time reversal R, can be replaced by a translation T about half the 
negative (x-y)-diagonal of the square unit-cell indicated by the arrow. 
So $2_z$T and RT are symmetry 
elements of the space group. Consequently, $2_z$ and R turn out to be 
symmetry elements of the corresponding magnetic point group obtained by 
setting T equal to unity. (Other symmetry operations, as for example 
reflection at the (x-z)- or (y-z)-plane, are not allowed for this 
configuration, since under these transformations single spins change sign, 
whereas others stay invariant. So the spin structure changes and 
cannot be restored by time reversal or a translation.) Dropping R in 
classifying a dynamical process the remaining point group is the classical 
monoclinic group $\mathbf{2}$ which consists of the elements 1 and $2_z$. 
 Hence, as can be seen from Tab. \ref{tab1},
 this spin configuration is especially suited for the detection of
 antiferromagnetism by nonlinear magneto-optics: 
The tensor 
elements $\mathit{xyz}$ = $\mathit{xzy}$, $\mathit{yzx}$ = $\mathit{yxz}$, 
and $\mathit{zxy}$ = $\mathit{zyx}$  appear in the antiferromagnetic 
phase only. Both in the para- and in the ferromagnetic state
($ \mathbf{M\parallel x} $) they are zero. On the other hand there are 
tensor elements, as for example $\mathit{yyy}$, $\mathit{yxx}$ and 
$\mathit{yzz}$, which occur exclusively in 
the ferromagnetic state.

 Having completed now the symmetry analysis of the 
susceptibility tensor the various tensor elements can be singled out in the 
SHG response by varying the light polarization ~\cite{bohm,hub}: The 
reflected light at 
frequency $ 2\omega $ is given by 
\begin{equation}
E^{(2\omega)}(\Phi,\phi)=2i\left( \frac{\omega}{c} \right)
|E^{(\omega)}_0|^2\left( \begin{array}{c} A_pF_c\cos\Phi \\
 A_s\sin\Phi \\ A_pN^2F_s\cos\Phi \end{array} \right)\chi^{(2\omega)}_{el} 
\end{equation} 
\[
\times\left( \begin{array}{c} f^2_ct^2_p\cos^2\varphi \\ t^2_s\sin^2\varphi \\
 f^2_st^2_p\cos^2\varphi \\ 2f_st_pt_s\cos\varphi\sin\varphi \\ 
2f_cf_st^2_p\cos^2\varphi \\ 2f_ct_pt_s\cos\varphi\sin\varphi 
\end{array} \right) ,
\]    
where $\chi^{(2\omega)}_{el}$ is the susceptibility tensor for 
the respective surface configuration, $\Phi,\varphi$ are the angles
 of polarization for the reflected second harmonic and fundamental light, 
$N(2\omega)$ is the index of refaction at frequency $2\omega$, 
$F_{c,s}
,f_{c,s}$ are the corresponding Fresnel coefficients, $T_{s,p},
t_{s,p}$ the linear transmission coefficients, and $A_p$, $A_s$ 
the amplitude factors of $\mathit{s}$- and $\mathit{p}$-polarized light. 
It becomes obvious how by 
varying the light polarization different elements of $\chi_{ijk}$ enter.
To discriminate optically 
the antiferromagnetic spin configuration of the (001) surface from the para- 
or ferromagnetic one it is sufficient to measure the $\mathit{s}$-polarized 
SHG output
 as a function of the polarization of the fundamental light. Tab. \ref{tab2} 
 shows the resulting SHG fields obtained by Eq. (4) with the 
characteristic tensor elements for the three phases and different incoming 
light polarizations. Obviously the 
difference between the polarized SHG signals of the states is caused 
by the appearance of the $\mathit{yyy}$ element in the ferromagnetic phase 
in contrast 
to the para- and antiferromagnetic ones 
($\mathit{s\rightarrow s}$) and by the appearance of the 
$\mathit{yzx}$ element in the antiferromagnetic state in contrast to 
the paramagnetic 
one ($\mathit{p\rightarrow s}$). Fig. \ref{fig2} schematically 
illustrates the differences between the phases in dependence on the 
polarization of incident fundamental and reflected SHG light: In the 
paramagnetic state the $\mathit{s}$-SHG signal vanishes for both
 $\mathit{s}$- and $\mathit{p}$-polarized incoming light, 
in the antiferromagnetic 
state it vanishes only 
for $\mathit{s}$- polarized incoming light and in the ferromagnetic state 
a pronounced 
s-SHG signal occurs for all three polarizations of the fundamental light. 
Thus, performing measurements with $\mathit{s}$-, $\mathit{p}$-, or 
mixed polarized 
incoming light allows to sort out the antiferromagnetic state unambiguously.
Moreover, using appropriate polarizations of incident and SHG light one 
can directly observe the transition between the antiferromagnetic and 
paramagnetic state upon varying the temperature: The 
characteristic SHG signal for the antiferromagnetic phase disappears upon 
crossing the N\'eel point. 

In conclusion, our analysis shows that nonlinear optics provides a 
very sensitive optical probe for antiferromagnetically ordered surfaces, 
even if the underlying threedimensional lattice is of very high symmetry. 
The time inversion symmetry of these surfaces needs not to be broken 
either ~\cite{erg}.
This could be important for the observation of surfaces of oxidized transition 
metals and oxides like NiO, which often exhibit the above mentioned 
symmetry. Our analysis of antiferromagnetic spin configurations can be 
extended to interfaces such as Fe/Cr and offers a possibility for the 
imaging of an ensemble of $\mathit{ferromagnetic}$ domains. 
An interesting application 
is also the determination of spin waves via time-resolved SHG spectroscopy. 

\end{multicols}

\newpage
\noindent
\begin{table}
\caption{\label{tab1}Nonvanishing elements of $\chi^{(2\omega)}_{el}$ 
for certain spin configurations of the (001), (110),and (111) surfaces 
of a fcc lattice. We denote $\chi_{ijk}$ by ijk.}
\begin{tabular}{cccll}
surface&configuration&point group&symmetry operations&non-vanishing 
independent tensor elements\\ \hline
 & & & & \\
(001)&para&4mm&$1,2_z,\pm4_z,\overline{2}_x,\overline{2}_y,\overline{2}_{xy},
\overline{2}_{-xy}$&$zxx = zyy,
 zzz, xxz = xzx = yyz = yzy$\\
  & & & & \\
$\prime\prime$&ferro$
$&-&$1,\underline{2}_z,
\overline{2}_x,\underline{\overline{2}}_y$&xxz
= xzx, zxx, yyz = yzy, zyy, zzz, \\
 &$(\mathbf{m\parallel x})$& & &zzy = zyz, yzz, xxy = xyx, yxx, yyy \\
 & & & & \\
$\prime\prime$&afm:& & & \\
 & a)&-&$1,\overline{2}_y$& xxx, yyx = yxy, xyy, xxz = xzx,
zxx, \\
 & & & & yyz = yzy, zzx = zxz, xzz,
 zyy, zzz \\
 &b)&-&$1,\overline{2}_x$& xxy = xyx, yxx, yyy, yzz, zzy = yzy \\ 
 & & & & xxz = xzx, zxx, yyz = yzy, zyy, zzz \\
 &c)&2&$1,2_z$&zxx, xxz = xzx, zyy, yyz = yzy, zzz, \\
 & & & &xyz = xzy, yzx = yxz,
zxy = zyx \\
 & & & &\\
(110)&para&mm2&$1,2_z,\overline{2}_x,\overline{2}_y$&zxx, xxz = xzx, 
zyy, yyz = yzy,
zzz \\
 & & & & \\
$\prime\prime$&afm&2&$1,2_z$&zxx, xxz = xzx, zyy, yyz = yzy, zzz, \\
 & & & & xyz = xzy, yzx = yxz,
zxy = zyx \\
 & & & & \\
(111)&para&3m&$1,\pm3_z,3(\overline{2}_{\perp})$&xxx = -xyy =
 -yyx = -yxy, zxx = zyy, \\
-more than one monolayer- & & & &xxz = xzx = yyz = yzy, zzz \\
$\prime\prime$&afm&-&$1,\overline{2}_y$&xxx, xyy, yyx = yxy, zxx, xxz = xzx, \\
 & & & &zyy, yyz = yzy, xzz, zzx = zxz, zzz \\
  & & & & \\
(111)&para&6mm&$1,2_z,\pm6_z,6(2_{\perp})$&zxx = zyy,
 xxz = xzx = yyz = yzy, zzz \\
-exactly one monolayer-& & & & \\
$\prime\prime$&afm&2&$1,2_z$&zxx, xxz = xzx, zyy, yyz = yzy, zzz, \\
 & & & &xyz = xzy, yzx = yxz, zxy = zyx \\
\end{tabular}
\end{table}

\begin{table}
\caption{\label{tab2}Reflected SHG signal for  different 
polarizations of the fundamental light for the (001) surface of a fcc crystal.}
\begin{tabular}{cccl}
configuration&$P_{incoming-light}$&$P_{SHG-light}$&SHG wave 
$E^{(2\omega)}$\\ \hline
para&s&s&0\\
    &p&s&0\\
    &mix($45^{\circ}$)&s& $ 2i\left(\frac{\omega}{c}\right)
|E^{(2\omega)}_0|^2A_sf_st_pt_sxxz $ \\
 & & &  \\
ferro&s&s&$ 2i\left(\frac{\omega}{c}\right)
|E^{(2\omega)}_0|^2A_
st^2_syyy $ \\
$(\mathbf{ M\parallel x })$&p&s& $ 2i\left(\frac{\omega}{c}\right)
|E^{(2\omega)}_0|^2
A_st^2_p(f^2_cyxx+f^2_syzz) $ \\
    &mix($45^{\circ}$)&s& $ 2i\left(\frac{\omega}{c}\right)
|E^{(2\omega)}_0|^2A_s(\frac{1}{2}f^2_cf^2_pyxx+\frac{1}{2}t^2_syyy+
\frac{1}{2}f^2_st^2_pyzz+f_st_pt _syyz) $ \\
 & & &  \\
afm&s&s&0\\
  &p&s& $ 2i\left(\frac{\omega}{c}\right)|E^{(2\omega)}_0|^2
A_st^2_p2f_cf_syzx $ \\
   &mix($45^{\circ}$)&s& $ 2i\left(\frac{\omega}{c}\right)
|E^{(2\omega)}_0|^2A_s(f_st_pt_syyz+f_cf_st^2_pyzx) $  
\end{tabular}
\end{table}
\newpage

\begin{figure}
\caption{\label{fig2}Dependence of the s-polarized SHG signal on the polarization of the incoming light.}
\end{figure}


\begin{thebibliography}{99}
\bibitem{bem} Note, for ferromagnets a similar interference of the odd 
and even elements of the electric susceptibility tensor
 $\chi^{(2\omega)}_{el}$ is required for the observation of the nonlinear 
Kerr effect.
\bibitem{zus} The complete symmetry classification will be given in a 
forthcoming publication.
\bibitem{nio}  Since bulk NiO has NaCl structure, i.e. it consists of 
two staggered fcc lattices, the unit-cell of the (001) surface is a square 
which contains two nickel ions carrying the spin moments and two 
oxygen ions.
\end{thebibliography}
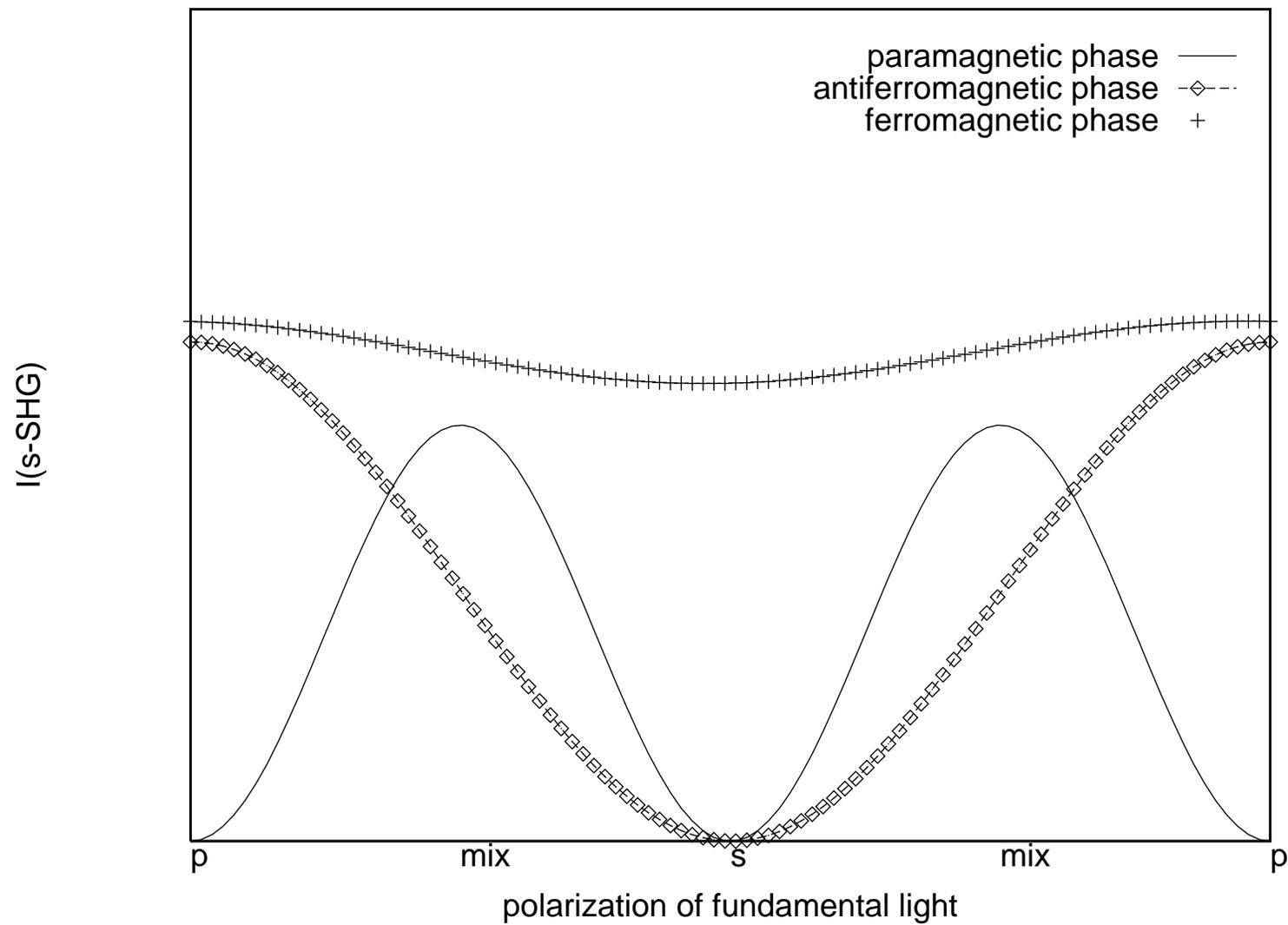
\end{document}